\documentclass[10pt]{article}
\date{}
\title{Additive Entropies of degree-$q$ and the 'Tsallis' Entropy}
   \author{B. H. Lavenda$^1$ and J. Dunning-Davies$^2$\\
$^1$Universit\`a degli Studi  Camerino 62032 (MC) Italy;\\ email: bernard.lavenda@unicam.it\\
$^2$ Department of Physics, University of Hull, Hull HU6 7RX\\
England; email: j.dunning-davies@hull.ac.uk}
\usepackage{eufrak}
\newcommand{\half}{\mbox{\small$\frac{1}{2}$}}

\newcommand{\fourthirds}{\mbox{\small$\frac{4}{3}$}}

\begin{document}
\maketitle
\begin{abstract}
The Tsallis entropy is shown to be an additive entropy of degree-q that 
information scientists have been using for almost forty years. Neither is 
it a unique solution to the nonadditive functional equation from which 
random entropies are derived. Notions of additivity, extensivity and 
homogeneity are clarified. The relation between mean code lengths in coding theory and 
various expressions for average entropies is discussed.\end{abstract}
\section{The \lq Tsallis\rq\ Entropy}
In 1988 Tsallis \cite{Tsallis}published a much quoted paper containing
an expression for the entropy which differed from the usual one used
in statistical mechanics. Previous to this, the R\'enyi
entropy was used as an interpolation formula that connected the
Hartley--Boltzmann entropy to the Shannon--Gibbs entropy.
Notwithstanding the fact that the R\'enyi entropy is additive, it
lacks many other properties that characterize the Shannon--Gibbs
entropy. For example, the R\'enyi entropy is not subadditive,
recursive, nor does it possess the branching and sum properties
\cite{Aczel}. The so-called Tsallis entropy fills this gap, while
being nonadditive, it has many other properties that resemble the
Shannon--Gibbs entropy. It is no wonder then that this entropy
fills an important gap.\par Yet, it appears odd, to say the least,
that information scientists have left such a gaping void in their
analysis of entropy functions. A closer analysis of the literature
reveals that this is not the case and, indeed, a normalized
Tsallis entropy seems to have first appeared in a 1967 paper by
Havrda and Charv\'at \cite{Havrda} who introduced the normalized
\lq Tsallis\rq\ entropy
\begin{equation}
S_{n,q}(p_1,\ldots,p_n)=\left(\sum_{i=1}^n\,p_i^q-1\right)\bigg/(2^{1-q}
-1) \label{eq:Tsallis}
\end{equation}
for a complete set of probabilities, $p_i$, i.e.
$\sum_{i=1}^np_i=1$, and parameter $q>0$, but $q\neq1$. The latter
requirement is necessary in order that (\ref{eq:Tsallis}) possess
the fundamental property of the entropy; that is, it is a concave
function. According to Tsallis \cite{Tsallis99}, only for $q>0$ is
the entropy, (\ref{eq:Tsallis}), said to be
\emph{expansible\/} \cite{Aczel} [cf. (\ref{eq:exp})
below].\par \section{Properties of Additive Entropy of Degree-$q$}
The properties used to characterize the entropy are
\cite{Aczel,Mathai}:
\begin{enumerate}
\item Concavity
\begin{equation}
 S_{n,q}\left(\sum_{i=1}^n\,p_ix_i\right)\ge\sum_{i=1}^n\,p_iS_{n,q}(x_i) 
\label{eq:concavity}
\end{equation}
where the nonnegative $n$-tuple, $(p)=(p_1,\ldots,p_n)$, forms a complete probability
distribution. For ordinary means, the $n$-tuple, $(x)=(x_1,\ldots,x_n)$, represents a set of
nonnegative numbers which constitute a set independent variables. What constitutes the 
main difficulty in proving theorems on characterizing entropy functions in information
theory is that the \lq independent variables\rq, $(x)$,  are not independent of their \lq weights\rq, $(p)$ \cite{Aczel66}.\par
 Coding theory, to be discussed in the next section, derives the functional dependencies
in a very elegant way  through optimization. The entropies
$S(x_i)$ represent the costs of encoding a sequence of lengths $x_i$, whose 
probabilities are $p_i$. Minimizing the mean length associated with the cost function, 
 expressed as the weighted mean of the cost function, 
gives the optimal codeword lengths $x_i$ as functions of 
their probabilities, $p_i$. Consequently, the entropies that result when the $x_i$ are
evaluated at their optimal values by expressing them in terms of their probabilities, 
$p_i$, constitute lower bounds to the mean lengths for the cost function.
\item Non-negativity
\begin{equation} S_{n,q}(p_1,\ldots, p_n)\ge0 \label{eq:nn}
\end{equation}
\item Symmetry
\begin{equation}
S_{n,q}(p_1,\ldots,p_n)=S_{n,q}(p_{[1]},\ldots,p_{[n]})
\label{eq:symmetry}
\end{equation}
where $[]$ denotes any arbitrary permutation of the indices on the
probabilities. For the entropy, the symmetry property
(\ref{eq:symmetry}) means that it should not depend upon the order
in which the outcomes are labelled. \item The sum property
\begin{equation}
S_{n,q}(p_1,\ldots,p_n)=\sum_{i=1}^n\mathfrak{S}_{n,q}(p_i)
\label{eq:sum}
\end{equation}
where $\mathfrak{S}_{n,q}$ is a measurable function on $]0,1[$.
\item Expansibility
\begin{equation}
S_{n+1,q}(0,p_1,\ldots,p_n)=S_{n,q}(p_1,\ldots,p_n),
\label{eq:exp}
\end{equation}
meaning that the entropy should not change when an outcome of
probability zero is added. \item Recursivity of degree-$q$
\begin{equation}
S_{n,q}(p_1,\ldots,p_n)=S_{n-1,q}(p_1+p_2,p_3,\ldots,p_n)+(p_1+p_2)^q
S_{2,q}\left(\frac{p_1}{p_1+p_2},\frac{p_2}{p_1+p_2}\right),
\label{eq:recur}
\end{equation}
asserting that if a choice is split into two successive choices,
the original entropy will be the weighted sum of the individual
entropies. Recursivity implies the \emph{branching\/} property by
requiring at the same time the additivity of the entropy as well
as the weighting of the different entropies by their corresponding
probabilities \cite{Renyi70}. \item Normality
\begin{equation}
S_{2,q}\left(\half,\half\right)=1 \label{eq:normal} \end{equation}
\item Decisivity
\begin{equation}
S_{2,q}(1,0)=S_{2,q}(0,1)=0 \label{eq:decis}
\end{equation}
\item Additivity of degree-$q$
\begin{eqnarray}
\lefteqn{S_{nm,q}(p_1q_1,\ldots,p_1q_m,\ldots,p_nq_m) =  S_{n,q}(p_1,\ldots,p_n)}
\label{eq:add}\\
& + & (2^{1-q}-1)
S_{n,q}(p_1,\ldots,p_n)S_{m,q}(q_1,\ldots,q_m)\nonumber
\end{eqnarray}
for any two complete sets of probabilities, $(p)$ and $(q)$. As
late as 1999, Tsallis \cite{Tsallis99} refers to (\ref{eq:add}) as
exhibiting \lq\lq a property which has apparently never been
focused before, and which we shall refer to as the
\emph{composability\/} property.\rq\rq\ Here, composability means
something different than in information theory \cite{Aczel}, in
that it \lq\lq concerns the nontrivial fact that the entropy
$S(A+B)$ of a system composed of two independent subsystems $A$
and $B$ can be calculated from the entropies $S(A)$ and $S(B)$ of
the subsystems, \emph{without any need of the microscopic
knowledge about $A$ and $B$, other than the knowledge of some
generic universality class, herein the nonextensive universality
class, represented by the entropic index $q\ldots$}\rq\rq
\cite{Tsallis99}.\par 
However, the additive entropy of degree-$q$, (\ref{eq:Tsallis}),  
 is not the only solution to the functional equation (\ref{eq:add}) for $q\neq1$. 
The average entropy
\begin{equation}
S^A_{n,q}(p_1,\ldots,p_n)=\frac{q}{q-1}\left\{1-\left(\sum_{i=1}^n p_i^q\right)^{1/q}  \right\}
\label{eq:Arimoto}
\end{equation}
also satisfies (\ref{eq:add}), with the only
difference that $(1-q)/q$ replaces the coefficient in the multiplicative term 
\cite{Lubbe}. Since the weighted mean of degree-$q$ is homogeneous, the 
pseudo-additive entropy
(\ref{eq:Arimoto}) is a first-order homogeneous function of $(p)$, $S^A_{n,q}
(\lambda p_1,
\ldots,\lambda p_n)=\lambda S^A_{n,q}(p_1,\ldots,p_n)$.  
It can be derived by averaging the same solution to the functional
equation (\ref{eq:add}), in the case $q\neq1$, as that used to derive the Tsallis 
entropy, except with a different exponent and normalizing factor, 
under the constraint that the probability distribution is complete \cite{Arimoto}. 
Although the pseudo-additive entropy (\ref{eq:Arimoto}) 
lacks the property of recursivity, (\ref{eq:recur}), it is monotonic, continuous, and 
concave for all positive values of $q$. Weighted means have been shown to be measures
of the extent of a distribution \cite{Campbell66}, and (\ref{eq:Arimoto}) relates the 
entropy to the weighted mean rather than to the more familiar logarithm of the weighted
mean, as in the case of the Shannon and R\'enyi entropies.
\par
Tsallis, in fact, associates additivity with
extensivity in the sense that for independent subsystems
\begin{equation}
S_{nm,q}(p_1q_1,\ldots,p_nq_m)=S_{n,q}(p_1,\ldots,p_n)+S_{m,q}(q_1,\ldots,q_m)
\label{eq:add-bis}
\end{equation}
According to Tsallis \cite{Tsallis99}, \emph{superadditivity\/},
$q<1$, would correspond to \emph{superextensivity\/}, and
\emph{subadditivity\/}, $q>1$, would correspond to
\emph{subextensivity\/}. According to Callen \cite{Callen},
extensive parameters have values in a composite system that are
equal to the sum of the values in each of the systems. Anything
that is not \emph{extensive\/} is labelled \emph{intensive\/},
although Tsallis would not agree [cf. (\ref{eq:beta}) below]. For
instance if we consider black-body radiation in a cavity of volume
$V$, having an internal energy, $U$, and magnify it $\lambda$
times, the resulting entropy
\begin{equation}\lambda S(U,V)=\fourthirds\sigma^{1/4}(\lambda
U)^{3/4}(\lambda V)^{1/4},\label{eq:bb} \end{equation}  will be
$\lambda$ times the original entropy, $S(U,V)$, where $\sigma$ is
the Stefan-Boltzmann constant. Whereas extensitivity involves
magnifying  \emph{all\/} the extensive variables by the same
proportion, additivity in the sense of being superadditive or
subadditive deals with a subclass of extensive variables, because
the condition of extensivity of the entropy imposes that the
determinant formed from the second derivatives of the entropy
vanish \cite{Lavenda}. The entropy of black-body radiation,
(\ref{eq:bb}), is extensive yet it is subadditive in either of the
extensive variables. The property of subadditivity is what Lorentz
used to show how interactions lead to a continual increase in
entropy \cite{Lavenda}. This is a simple consequence of
Minkowski's inequality,
\[u_1^{3/4}+u_2^{3/4}\ge (u_1+u_2)^{3/4},\] where $u=U/V$ is the
energy density. Hence, (sub-or super-) extensivity is something
very different from (sub-or super-) additivity. \item Strong
additivity of degree-$q$
\begin{eqnarray}
\lefteqn{S_{mn,q}(p_1q_{11},\ldots,p_nq_{1m},\ldots,p_nq_{nm})=S_{n,q}(p_1,\ldots,p_n)}\nonumber\\
& + & \sum_{j=1}^np_j^qS_{m,q}(q_{j1},\ldots,q_{jm})
\label{eq:strong}
\end{eqnarray}
where $q_{ij}$ is the conditional probability. Strong additivity
of degree-$q$ describes the situation in which the sets of
outcomes of two experiments are not independent. Additivity of
degree-$q$, (\ref{eq:add}),  follows from strong additivity by
setting $q_{1k}=q_{2k}=\cdots=q_{mk}=q_k$, and taking
(\ref{eq:Tsallis}) into consideration \cite{Aczel}. \par A doubly
stochastic matrix $(q_{ij})$, where $m=n$, is used in
\emph{majorization\/} to distribute things, like income, more
evenly \cite{Marshall}, and this leads to an increase in entropy.
For if
\begin{equation}q_j=\sum_{i=1}^nq_{ij}p_i,\label{eq:major}
\end{equation}
 and
\[\sum_{j=1}^nq_j=\sum_{i=1}^np_i\sum_{j=1}^nq_{ij}=\sum_{i=1}^np_i=1,\]
it follows from the convexity of $\psi=x\ln x$, or
\[\psi\left(\sum_{i=1}^nq_{ij}p_i\right)\le\sum_{i=1}^nq_{ij}\psi(p_i),\]
that
\begin{equation}S_{n,1}(q_1,\ldots,q_n)=-\sum_{i=1}^nq_i\ln
q_i\ge-\sum_{i=1}^n\sum_{j=1}^nq_{ij}p_i\ln p_i\label{eq:Shannon}
\end{equation}
since $\sum_{i=1}^nq_{ij}=1$. We may say that $p$ majorizes $q$,
$p\succ q$ if and only if (\ref{eq:major}) for some doubly
stochastic matrix $(q_{ij})$ \cite{Hardy}. A more even spread of
incomes increases the entropy. Here we are at the limits of
equilibrium thermodynamics because we are invoking a mechanism for
the increase in entropy, which in the case of incomes means taking
from the rich and giving to the poor \cite{Arnold}. This restricts
$q$ in the \lq Tsallis\rq\ entropy to  $]0,1[$. Values of $q$ in
$]1,2[$ show an opposing tendency of balayage or sweeping out
\cite{Lavenda03}. Whereas averaging tends to decrease inequality,
balayage tends to increase it \cite{Arnold}.\par Yet Tsallis
\cite{Tsallis99} refers to processes with $q<1$, i.e. $p^q_i>p_i$,
as \emph{rare\/} events, and to $q>1$, i.e. $p_i^q<p_i$ as
\emph{frequent\/} events. However, only in the case where $q<1$
will the Shannon entropy, (\ref{eq:Shannon}) be a lower bound to
other entropies like, the R\'enyi entropy
\begin{equation}
S_{n,q}^R=\frac{1}{1-q}\left(\ln\sum_{i=1}^np_i^q\right)
\label{eq:Renyi}
\end{equation}
which is the negative logarithm of the weighted mean of
$p^{q-1}_i$. The R\'enyi entropy has the attributes of reducing to
the Shannon--Gibbs entropy, (\ref{eq:Shannon}), in the limit as
$q\rightarrow1$, and to the Hartley--Boltzmann, entropy
\begin{equation}
S_{n,0}(1/n,\ldots,1/n)=\ln n \label{eq:Hartley}
\end{equation}
in the case of equal \emph{a priori\/} probabilities $p_i=1/n$.
This leads to the property of \item $n$-maximal
\begin{equation}
S_{n,q}(p_1,\ldots,p_n)\le
S_{n,q}\left(\frac{1}{n},\ldots\frac{1}{n}\right) \label{eq:mono}
\end{equation}
for any given integer $n\ge2$. The right-hand side of
(\ref{eq:mono}) should be a monotonic increasing function of $n$.
As we have seen, the tendency of the entropy to increase as the
distribution becomes more uniform is due to the property of
concavity (\ref{eq:concavity}). Hence, it would appear that
processes with $q<1$ would be compatible with the second law of
thermodynamics, rather than being \emph{rare\/} exceptions to
it!\item Continuity: The entropy is a continuous function of its
$n$ variables. Small changes in the probability cause
correspondingly small changes in the entropy. Additive entropies
of degree-$q$ are small for small probabilities, i.e.,
\[\lim_{p\rightarrow0}S_{2,q}(p)=\lim_{p\rightarrow0}\frac{p^q+(1-p)^q-1}{2^{1-q}-1}.\]
\end{enumerate}
\section{Coding Theory and Entropy Functions}
The analogy between coding theory and entropy functions has long
been known \cite{Campbell}. If $k_1,\ldots,k_n$ are the lengths of
codewords of a uniquely decipherable code with $D$ symbols then
the average codeword length \begin{equation}
\sum_{i=1}^np_ik_i\label{eq:S-average} \end{equation}
 is bounded
from below by the Shannon-Gibbs entropy (\ref{eq:Shannon}) if the
logarithm is to the base $D$. The optimal codeword length is
$k_i=-\ln p_i$, which represents the information content in event
$E_i$. If $D=\half$ then $p_i=\half$ contains exactly one bit of
information.\par Ordinarily, one tries to keep the average
codeword length (\ref{eq:S-average}) small, but it cannot be made
smaller than the Shannon-Gibbs entropy. An economical code has
frequently occurring messages with large $p_i$ and small $k_i$.
Rare messages are those with small $p_i$ and large $k_i$. The
solution $n_i=-\ln p_i$ has the disadvantage that the codeword
length is very great if the probability of the symbol is very
small. A better measure of the codeword length would be
\begin{equation}
\frac{1}{\tau}\log\left(\sum_{i=1}^np_iD^{\tau k_i}\right)
\label{eq:R-average} \end{equation} where $\tau=(1-q)/q$, thereby
limiting $q$ to the interval $[0,1]$. As $\tau\rightarrow\infty$,
the limit of (\ref{eq:R-average}) is the largest of the $k_i$,
independent of $p_i$. Therefore,
if $q$ is small enough, or $\tau$ large enough, the very large
$k_i$'s will contribute very strongly to the average codeword
length (\ref{eq:R-average}), thus keeping it from being small even
for very small $p_i$. The optimal codeword length is now
\[k_i=-q\ln p_i+\sum_{i=1}^np_i^q,\]
showing that the R\'enyi entropy is the lower bound to the average
codeword length (\ref{eq:R-average}) \cite{Campbell}. Just as the
 $p_i=D^{-k_i}$ are the optimum probabilities for the
 Shannon-Gibbs
 entropy, the optimum probabilities for the R\'enyi entropy are
 the so-called \emph{escort\/} probabilities,
 \begin{equation}
 D^{-k_i}=\frac{p_i^q}{\sum_{i=1}^np_i^q}.\label{eq:escort}
 \end{equation} As $p_i\rightarrow0$, the
 optimum value of $k_i$ is asymptotic to $-q\ln p_i$ so that
 the optimum length is less than $-\ln p_i$ for $q<1$ and sufficiently small $p_i$.
 This provides additional support for keeping $q$ within the interval $[0,1]$ \cite{Lavenda03}.\par
Although the R\'enyi entropy is additive it does not have other
properties listed above; for instance, it is not recursive and
does not have the branching property nor the sum property. It is
precisely the \lq Tsallis\rq\ entropy which fills the gap, while
not being additive, it has many of the other properties that an
entropy should have \cite{Daroczy}. Therefore, in many ways the
additive entropy of degree-$q$ (\ref{eq:Tsallis}) is closer to the
Shannon entropy, (\ref{eq:Shannon}) than the R\'enyi entropy is.
The so-called additive entropies of degree-$q$  can be written as
\begin{equation}
S_{n,q}(p_1,\ldots,p_n)=\sum_{i=2}^n(p_1+\ldots+p_i)^qf\left(\frac{p_i}{p_1+\ldots+p_i}\right),
\label{Curado}
\end{equation}
where the function $f$ is a solution to the functional equation
\[f(x)+(1-x)^qf\left(\frac{y}{1-x}\right)=f(y)+(1-y)^qf\left(\frac{x}{1-y}\right),\]
subject to $f(0)=f(1)$, which was rederived by Curado and Tsallis
\cite{Curado}, and the property of additivity of degree-$q$
(\ref{eq:add}) was referred to them as \emph{pseudo-additivity\/},
omitting the original references. What these authors appeared to
have missed  are the properties of strong additivity,
(\ref{eq:strong}) and recursivity of degree-$q$ (\ref{eq:recur}).
These properties can be proven by direct calculation using the
normalized additive entropy of degree-$q$, (\ref{eq:Tsallis}).
Additive entropies of degree-$q\ge1$ are also subadditive.\par
Moreover, additive entropies of degree-$q$ satisfy the sum
property, (\ref{eq:sum}) where
\begin{equation}
\mathfrak{S}_q(p_i)=(p_i^q-p_i)/(2^{1-q}-1)\ge0.
\label{eq:sum-bis}
\end{equation}
Only for $q>0$ will (\ref{eq:sum-bis}), and consequently
(\ref{eq:Tsallis}), be concave since
\[\mathfrak{S}^{\prime\prime}_q(p_i)=q(q-1)p_i^{q-2}/(2^{1-q}-1)\le0,\]
where the prime stands for differentiation with respect to $p_i$.
This is contrary to the claim that the additive entropy of
degree-$q$ is \lq\lq extremized \emph{for all values of
$q$\/}\rq\rq\cite{Tsallis}. It can easily be shown that the
concavity property
\[\mathfrak{S}_q\left(\sum_{i=1}^np_ix_i\right)\ge\sum_{i=1}^np_i
\mathfrak{S}_q(x_i),\] implies the monotonic increase in the
entropy (\ref{eq:mono}). Setting $p_i=1/n$ and using the sum
property (\ref{eq:sum}) lead to
\[S_{n,q}(p_1,\ldots,p_n)=\sum_{i=1}^n\mathfrak{S}_q(p_i)\le
n\mathfrak{S}_q\sum_{i=1}^n\left(\frac{p_i}{n}\right)=n\mathfrak{S}_q\left(
\frac{1}{n}\right)=S_{n,q}\left(\frac{1}{n},\ldots,\frac{1}{n}\right),\]
showing that $S_{n,q}(1/n,\ldots,1/n)$ is maximal.\par In order to
obtain explicit expressions for the probabilities, Tsallis
and collaborators maximized their non-normalized entropy
\begin{equation}
S^T_{n,q}(p_1\ldots,p_n)=\left(\sum_{i=1}^np^q_i-1\right)/(1-q)
\label{eq:Tsallis-bis}
\end{equation}
with respect to certain constraints. Taking their cue from Jaynes'
\cite{Jaynes} formalism of maximum entropy, (\ref{eq:Tsallis-bis})
was to be maximized with respect to the finite norm
\cite{Tsallis95}
\[
\int_{-\infty}^{\infty}\,p(x)\,dx=1\] and the so-called $q$
average of the second moment \cite{Curado}
\begin{equation}
\left<x^2\right>_{q}=\int_{-\infty}^{\infty}x^2
[\sigma\,p(x)]^{q}\,d(x/\sigma)=\sigma^2.\label{eq:2nd}
\end{equation}
The latter condition was introduced because the variance of the
distribution did not exist, and the weights, $(p^q)$, have been
referred to as \lq escort\rq\ probabilities [cf. (\ref{eq:escort})
above]. The resulting distribution is almost identical to
Student's distribution
\begin{equation}
p(x|\mu)=\sqrt{\frac{\mu}{\pi}}\frac{(q-1)}{(3-q)}\frac{\Gamma\left(1/(q-1)\right)}
{\Gamma\left((3-q)/2(q-1)\right)}\left(1+\frac{(q-1)}{(3-q)}\mu
x^2\right)^{-1/(q-1)} \label{eq:Student}
\end{equation}
where  $(3-q)/(q-1)$ is the number of degrees of freedom, and
$\mu$ is the Lagrange multiplier for the constraint
(\ref{eq:2nd}) \cite{Souza}.\par The Gaussian distribution is the only stable
law with a finite variance, all the other stable laws have
infinite variance. These stable laws have much larger tails than
the normal law which is responsible for the infinite nature of
their variances. Their initial distributions are given by the
intensity of small jumps, where the intensity of jumps having the
same sign of $x$, and greater than $x$ in absolute value is
\cite{deFinetti}
\begin{equation}
\overline{F}(x)=\frac{c}{x^{\beta}},\label{eq:F}
\end{equation}
for $x>1$. For $\beta<1$, the generalized random process, which is
of a Poisson nature, produces only positive jumps, whose intensity
(\ref{eq:F}) is always increasing. No moments exist, and the fact
that
\begin{equation}
\mathcal{Z}(\lambda)=e^{-\lambda^{\beta}} \label{eq:Z}
\end{equation}
where $\lambda$ is both positive and real, follows directly from
P\'olya's theorem: If for each $\lambda$, $\mathcal{Z}(0)=1$,
$\mathcal{Z}(\lambda)\ge0$,
$\mathcal{Z}(\lambda)=\mathcal{Z}(-\lambda)$,
$\mathcal{Z}(\lambda)$ is decreasing and continuous convex on the
right half interval, then $Z(\lambda)$ is a generating function
\cite{Chung}. Convexity is easily checked for $0<\beta\le1$, and
it is concluded that $z(\lambda)$ is a generating function. In
other words,
\[
1-\mathcal{Z}(\lambda)=-\int_{0}^{\infty}\left(1-e^{-\lambda
x}\right)\,d\overline{F}(x)=\Gamma(1-\beta)\lambda^{-\beta}
\]exists for a positive argument of the Gamma function, and that
implies  $\beta<1$. \par This does not hold on the interval
$1<\beta<2$, where it makes sense to talk about a compensated sum
of jumps, since a finite mean exists. In the limit $\beta=2$,
positive and negative jumps about the mean value become equally as
probable and the Wiener-L\'evy process results, which is the
normal limit. If one introduces a centering term in the
expression, $\lambda x$, the same expression for the generating
function, (\ref{eq:Z}), is obtained to lowest power in $\lambda$,
as $\lambda\rightarrow0$ and $x\rightarrow\infty$, such that their
product is finite.
\par
These stable distributions, $0<\beta<1$, (and quasi-stable ones,
$1<\beta<2$, because the effect of partial compensation of jumps
introduces an arbitrary additive constant)  are related to the
process of super-diffusion, where the asymptotic behavior of the
generalized Poisson process has independent increments with
intensity (\ref{eq:F}). For strictly stable processes, the
super-diffusion packet spreads out faster than the packet of
freely moving particles, while  a quasi-stable distribution
describes the random walk of a particle with a finite mean
velocity. It was hoped that these tail distributions could be
described by an additive entropy of degree-$q$, where the degree
of additivity would be related to the exponent of the stable, or
quasi-stable, distribution. Following the lead of maximum entropy,
where the optimal distribution results from maximizing the entropy
with all that is known about the system, the same would hold true
for maximizing the additive entropy of degree-$q$. However, it was
immediately realized that the variance of the distribution does
not exist.\par Comparing the derivative of the tail density
(\ref{eq:F}) with (\ref{eq:Student}) identifies
$\beta=(3-q)/(q-1)$, requiring the stable laws to fall in the
domain $\mbox{\small{$\frac{5}{3}$}}<q<3$ \cite{Tsallis95}.
However, it is precisely in the case in which we are ignorant of
the variance that the Student distribution is used to replace the
normal since it has much fatter tails and only approaches the
latter as the number of degrees of freedom increases without limit
\cite{deFinetti}. Just as the ratio of the difference of the mean
of a sample and the mean of the distribution to the standard
deviation is distributed normally, the replacement of the standard
deviation by its estimator is distributed according to the
Student's distribution. This distribution (\ref{eq:Student}) was
not to be unexpected,  because it stands in the same relation to
the normal law as the  \lq Tsallis\rq\ entropy,
(\ref{eq:Tsallis-bis}), in the limit as the number of degrees of
freedom is allowed to increase without limit. \par Whereas
weighted means of order-$q$
\[\mathcal{M}_q=\left(\frac{\sum_{i=1}^np_ix_i^q}{\sum_{i=1}^np_i}\right)^{1/q}\]
do have physical relevance for different values of $q$, the
so-called $q$-expectation
\[\left<x\right>_q:=\frac{\sum_{i=1}^np_i^qx_i}{\sum_{i=1}^np^q_i},\]
has no physical significance for values of $q\neq1$. Since the
connection between statistical mechanics and thermodynamics lies
in the association of average values with thermodynamic variables,
the $q$-expectations would lead to incorrect averages. This
explains why for Tsallis the internal energy of a composite system
is not the same as the internal energies of the subsystems, and
makes the question \lq\lq if we are willing to consider the
nonadditivity of the entropy, why it is so strange to accept the
same for the energy?\rq\rq \cite{Tsallis98} completely
meaningless. Yet, the zeroth law of thermodynamics, and the
derivation of the Tsallis nonintensive inverse temperature,
\begin{equation}\beta=\frac{\partial S_{n,q}}{\partial
U_q}\bigg/[1-(1-q)S_{n,q}],\label{eq:beta} \end{equation} where
$U_q$ is the $q$-expectation of the internal energy, rest on the
fact that the total energy of the composite system is conserved
\cite{Abe}.\par It is as incorrect to speak of \lq Tsallis\rq\
statistics \cite{Science} as it would be to talk of R\'enyi
statistics. These expressions  are mere interpolation formulas
leading to statistically meaningful expressions for the entropy in
certain well-defined limits. Whereas for the R\'enyi entropy the
limits $q\rightarrow1$ and $q\rightarrow0$ give the Shannon-Gibbs
and Hartley-Boltzmann entropies, respectively, without assuming
equal probabilities, the additive entropy of degree-$q$ reduces to
the Shannon entropy in the limit as $q\rightarrow1$, but it must
further be assumed that the \emph{a priori\/} probabilities are
equal in order to reduce it to the Hartley-Boltzmann entropy.
Hence, only the R\'enyi entropies are true interpolation formulas.
\par Either the average of $-\ln p_i$ leading to the Shannon
entropy, or the negative of the weighted average of $p_i^{q-1}$,
resulting in the R\'enyi entropy will give the property of
additivity \cite{Aczel}. Whereas the Shannon entropy is the
negative of the logarithm of the geometric mean of the
probabilities,
\[S_{n,1}(p_1,\ldots,p_n)=-\ln\mathfrak{G}_n(p_1,\ldots,p_n),\]
where
\[\mathfrak{G}_n(p_1\ldots,p_n)=\Pi_{i=1}^np_i^{p_i}\]
is the geometric mean, the R\'enyi entropy is the negative of the
logarithm of the weighted mean
\[S_{n,q}^R=-\ln\mathcal{M}_{q-1},\]
where
\[\mathcal{M}_{q-1}=\left(\sum_{i=1}^np_ip_i^{q-1}\right)^{1/(q-1)}\]
is the weighted mean of $p_i^{q-1}$. If the logarithm is to the
base 2, the additive entropies of degree-$q$ are exponentially
related to the R\'enyi entropies of order-$q$ by
\[S_{n,q}=\left(2^{(1-q)S_{n,q}^R}-1\right)\big/(2^{1-q}-1),\]
which make it apparent that they cannot be additive. But
nonadditivity has nothing to do with  nonextensivity.\par As a
concluding remark it may be of interest to note that undoubtedly
the oldest expression for an additive entropy of degree-2 was
introduced by Gini \cite{Gini} in 1912, who used it as an index of
diversity or inequality. Moreover, generalizations of additive
entropies of degree-$q$ are well-known. It has been claimed that
\lq\lq Tsallis changed the mathematical form of the definition of
entropy and introduced a new parameter $q$\rq\rq \cite{Chou}.
Generalizations that introduce  additive entropies of
degree-$q+r_i-1$ \cite{Rathie}
\[S_{n,q,r_1,\ldots,r_n}(p_1,\ldots,p_n)=\left(\frac{\sum_{i=1}^np_i^{q+r_i-1}}{
\sum_{i=1}^np_i^{r_i}}-1\right)\bigg/(2^{1-q}-1),\] with $n+1$
parameters, should give even better results when it comes to curve
fitting.

\end{document}